\documentstyle[12pt,epsf]{article}
\def\singlespace 
{\smallskipamount=3.75pt plus1pt minus1pt
\medskipamount=7.5pt plus2pt minus2pt
\bigskipamount=15pa plus4pa minus4pt \normalbaselineskip=12pt plus0pt
minus0pt \normallineskip=1pt \normallineskiplimit=0pt \jot=3.75pt
{\def\smallskip {\vskip\smallskipamount}} {\def\medskip
{\vskip\medskipamount}} {\def\bigskip {\vskip\bigskipamount}}
{\setbox\strutbox=\hbox{\vrule height10.5pt depth4.5pt width 0pt}}
\parskip 7.5pt \normalbaselines} 

\def\middlespace
{\smallskipamount=5.625pt plus1.5pt minus1.5pt \medskipamount=11.25pt
plus3pt minus3pt \bigskipamount=22.5pt plus6pt minus6pt
\normalbaselineskip=22.5pt plus0pt minus0pt \normallineskip=1pt
\normallineskiplimit=0pt \jot=5.625pt {\def\smallskip
{\vskip\smallskipamount}} {\def\medskip {\vskip\medskipamount}}
{\def\bigskip {\vskip\bigskipamount}} {\setbox\strutbox=\hbox{\vrule
height15.75pt depth6.75pt width 0pt}} \parskip 11.25pt
\normalbaselines} 

\def\doublespace 
{\smallskipamount=7.5pt plus2pt minus2pt \medskipamount=15pt plus4pt
minus4pt \bigskipamount=30pt plus8pt minus8pt \normalbaselineskip=30pt
plus0pt minus0pt
\normallineskip=2pt \normallineskiplimit=0pt \jot=7.5pt
{\def\smallskip {\vskip\smallskipamount}} {\def\medskip
{\vskip\medskipamount}} {\def\bigskip {\vskip\bigskipamount}}
{\setbox\strutbox=\hbox{\vrule height21.0pt depth9.0pt width 0pt}}
\parskip 15.0pt \normalbaselines}

\newcommand{\be}{\begin{eqnarray}}
\newcommand{\ee}{\end{eqnarray}}
\newcommand{\etal}{{\it et al.}}

\def\sst{{s_\theta}}
\def\ss2t{{s_{2 \theta}}}
\def\cct{{c_\theta}}
\def\cc2t{{c_{2 \theta}}}

\newcommand{\memu}{\mbox{$m_{e\mu}$~}}
\newcommand{\metau}{\mbox{$m_{e\tau}$~}}
\newcommand{\mmutau}{\mbox{$m_{\mu\tau}$~}}

\begin{document}
\middlespace

\vskip 2cm
\begin{flushright} SINP/TNP/01-25\\
\underbar{Accepted in Physics Letters B}
\end{flushright}
\begin{center}
\Large {\bf Viability of bimaximal solution of the Zee mass matrix} \\ 
\vskip 1cm Biswajoy Brahmachari 
\footnote{{\tt electronic address:biswajoy@theory.saha.ernet.in}}
and Sandhya Choubey 
\footnote{{\tt electronic address:sandhya@theory.saha.ernet.in}}\\
\end{center}
\begin{center}
Theoretical Physics Group\\
Saha Institute of Nuclear Physics\\
1/AF Bidhannagar, Kolkata-700064, India
\\

\end{center}
\vskip 2cm 
{
\begin{center}
\underbar{Abstract} \\
\end{center}

We know $L_e-L_\mu-L_\tau$ symmetry gives $m^2_1= m^2_2 >> m^2_3$ 
pattern in Zee model.  $\Delta m^2_\odot$ emerges from a small breaking
of this symmetry. Because this symmetry is broken very weakly  $\theta_\odot$ 
does not deviate much from $\tan^2 \theta_\odot=1$ which is its 
value in the symmetric limit. This gives a mismatch with LMA solution 
where mixing is large but not exactly maximal. We confront this property of 
Zee mass matrix by phenomenologically analyzing recent results from solar 
and atmospheric neutrino oscillation experiments at various confidence levels. 
We conclude that LOW type solution is compatible with the Zee mass matrix
at 99\% confidence level when atmospheric neutrino deficit is explained by 
maximal $\nu_\mu \leftrightarrow \nu_\tau$ oscillation. 
Thus the minimal version of the Zee model even though disfavored by the 
LMA type or VO type solutions, is compatible with LOW type solution of 
solar neutrino problem. 

\newpage

The neutrino mass matrix under the Zee ansatz \cite{zee} can be written in the 
flavor basis as 
\be
{\cal M} &=& {\pmatrix {0 & \memu & \metau \cr
                     \memu & 0   & \mmutau \cr
                     \metau & \mmutau & 0 \cr}}
\label{zee}
\ee
where
\be
m_{\alpha\beta} &=& (1/M)~f_{\alpha\beta}(m_{\beta}^2 - m_{\alpha}^2)\frac{
v_2}{v_1}
\ee
where $m_\alpha$ $(\alpha=e,\mu,\tau)$ are the masses of the charged leptons 
and $v_{1(2)}$ is the VEV of the neutral component of the two Higgs 
doublets $\Phi_{1(2)}$ required to complete the 
coupling $\Phi_1 \Phi_2 \chi$ where $\chi$ is the Zee
singlet which also couples to lepton doublets
via $f_{\alpha \beta} L_\alpha L_\beta \chi$ and $M$ is a 
mass parameter. The mass matrix (\ref{zee}) 
is symmetric because of its Majorana nature, off-dioganal because of the 
antisymmetry in $f_{\alpha \beta}$ and is real in three generations. 
There are two variations obtained from Eqn. (\ref{zee}). The first
one is due to Smirnov et. al. \cite{smirnov}. In this case one of the mass
squared difference is compatible with LSND data and the 
atmospheric neutrino problem is explained by maximal
$\nu_\mu \leftrightarrow \nu_\tau$
oscillations. As the mass of $\nu_\mu$ and $\nu_\tau$ lies in the
1 eV range they can form hot component of dark matter.
The second variation is of our interest \cite{jarlskog}. In this case
maximal $\nu_e \leftrightarrow \nu_\mu$ oscillation
leads to the solar neutrino deficit whereas maximal
$\nu_\mu \leftrightarrow \nu_\tau$ oscillations
lead to atmospheric neutrino deficit. It can be easily
seen that an approximate $L_e - L_\mu -L_\tau$ symmetry 
\cite{petcov} imposed
on the matrix in Eqn. (\ref{zee}) achieves the
goal \cite{jarlskog, frampton}. A large number of studies of the
Zee model exists in literature \cite{newpapers}.  
 
Even though the solar neutrino problem is best explained by invoking large 
mixing angles for the neutrinos, maximal mixing is 
disfavored in the LMA region. Hence Zee model runs into trouble 
since it predicts almost maximal mixing for the solar neutrinos
even if we allow for modest breaking of the $L_e-L_\mu-L_\tau$
symmetry to generate correctly 
the mass splittings 
$\Delta m^2_\odot$
needed for the depletion of
the solar neutrino flux. 
Now let us give some recent studies of Zee model
which will highlight the significance of this paper.
In Ref \cite{koide} it has been argued that Zee model
is in poor agreement with experimental data and thus
modifications of Zee model is necessary and some promising
modifications are also suggested. In Ref \cite{frampton1} it has
been argued that Zee model predicts maximal mixing solution
of the solar neutrino problem which is incompatible with
experimental data and two extensions of Zee model are proposed
which can accommodate the data. In this paper we take a closer look at
various regions where  large or maximal mixing solution of the solar
neutrino problem are allowed and confirm whether we need to go beyond the
minimal Zee model to accommodate present experimental data. The philosophy
behind our approach is that because Zee model is
very rich in physics and also quite predictive,
modification of the minimal version may become less
attractive. To do that we separately analyze the predictions 
of the Zee model in  three zones where large mixing angles 
are allowed from the solar neutrino problem.  They
are the large mixing angle (LMA) region with $\Delta m_{\odot}^2$ around
$5 \times 10^{-5}$ $eV^2$, the low $\Delta m^2$ (LOW) 
region with $\Delta m_{\odot}^2$
around $1 \times 10^{-7}$ $eV^2$ and the vacuum oscillation 
(VO) region with $\Delta m_{\odot}^2$ around $5 \times 10^{-10}$ $eV^2$. 
We will compare the prediction of the Zee model with the data in 
these three zones at various confidence levels and check the 
viability of the model. We will observe that the minimal Zee model is 
consistent with the experimental data at 99\% C.L.  
in the LOW region. 

We can re-express (\ref{zee}) in terms of parameters $m_0$, $\theta$ and 
$\epsilon$ 
in such a way that $m_0\sin\theta = {\cal M}_{e\mu}$, 
$m_0\cos\theta= {\cal M}_{e\tau}$ 
and $m_0\epsilon = {\cal M}_{\mu\tau}$. Then we get 
\begin{equation}
\tan\theta = \frac{f_{e\mu}}{f_{e\tau}}
\left(\frac{m_\mu^2}{m_\tau^2}\right)~~;~~
\epsilon = \frac{f_{\mu\tau}}{f_{e\tau}}\cos\theta
\end{equation}
The mass matrix then assumes the form,
\be
{\cal M} &=& m_0 {\pmatrix {0 & \sin\theta & \cos\theta \cr
                     \sin\theta & 0   & \epsilon \cr
                     \cos\theta & \epsilon & 0 \cr}}
\label{mass}
\ee
We will assume that the strengths of the coupling constants are 
such that $\epsilon \rightarrow 0$. Then we have a 
$L_e-L_\mu-L_\tau$ symmetry which is broken via
$M^\prime$ in the following notation, 
\be
{\cal M} &=&  m_0~{\pmatrix {0 & \sin\theta & \cos\theta \cr
                     \sin\theta & 0   & 0 \cr
                     \cos\theta & 0 & 0 \cr}} + 
               m_0~{\pmatrix {0 & 0 & 0 \cr
                    0  & 0   & \epsilon \cr
                     0 & \epsilon & 0 \cr}}  \\
&=& M_0 + M^\prime
\label{splitmass}
\ee
Now we can handle the diagonalization of the mass matrix (\ref{splitmass}) 
perturbatively, treating the $\epsilon$ as a small perturbation over 
$\theta$. With $\epsilon$ exactly zero the mass eigenvalues are 
\be
m^2_{1,2} = m_0^2~~;~~m^2_3=0,
\ee
while the mixing matrix which diagonalises $M_0$ is given by
\be
U = \frac{1}{\sqrt{2}}{\pmatrix {-1 & 1 & 0 \cr
               \sin\theta  & \sin\theta & \sqrt{2}\cos\theta \cr
               \cos\theta  & \cos\theta & -\sqrt{2}\sin\theta\cr}}
\label{unpermix}
\ee
We next impose the $\epsilon$ correction perturbatively and consider 
the first order corrections to the mass eigenvalues and the 
mixing matrix. The degeneracy between the $\nu_1$ and $\nu_2$ 
states are broken by the introduction of $\epsilon$ and the 
neutrino masses become,
\be
m_3 = -m_0\epsilon\sin 2\theta,~~m_{1,2} = m_0(\pm1+\frac{1}{2} \epsilon\sin 2\theta)
\label{m123}
\ee
So that the mass square differences become
\be
\Delta m_{13}^2 &\approx& \Delta m_{23}^2 = \Delta m_{atm}^2=m_0^2,\\ 
\Delta m_{12}^2 &=& \Delta m_{\odot}^2 = 2m_0^2\epsilon \sin 2\theta
\label{delmsq}
\ee
We define $s_x=\sin(x)$ and $c_x=\cos(x)$. The mixing matrix with 
the first order 
corrections assumes the form
\be
U = \frac{1}{\sqrt{2}}
{\pmatrix {
-1-\epsilon~ \ss2t/4
& 1- \epsilon~ \ss2t/4
& -\sqrt{2} \epsilon~ \cc2t \cr
\sst - \epsilon ~\sst \ss2t/4 - \epsilon ~\cct \cc2t
&
\sst + \epsilon ~\sst \ss2t/4 + \epsilon ~\cct \cc2t
& \sqrt{2} \cct \cr
\cct - \epsilon~ \cct \ss2t/4 + \epsilon ~\sst \cc2t
&
\cct + \epsilon ~\cct \ss2t/4 - \epsilon ~\sst \cc2t
& -\sqrt{2} \sst
}}
\label{permix}
\ee
At this stage the predictability of Zee model is clear.  For a given 
$\Delta m_{atm}^2$ and $\sin2\theta_{atm}$ (from the form 
of the mixing matrix (\ref{permix}) we see 
that $\theta\equiv \theta_{atm}$) and
for a given value of $\Delta m_{\odot}^2$ allowed by experimental data
at a certain confidence level, we can calculate the value of $\epsilon$
(See Fig \ref{fig1}).
Then using $\epsilon$ we can calculate three quantities from the 
Zee mass matrix, $\tan^2 \theta_\odot$, $\tan^2 \theta_{13}$ and 
$\Delta m^2_{13}$ and test the compatibility of Zee mass matrix with 
experimental data at that confidence level. This is what we propose to 
do in this paper. 

We begin by observing that the $\Delta m^2_{13}$ in this 
mass model is in the sensitivity range of the CHOOZ reactor 
experiment \cite{chooz}. If we use the standard Maki-Nakagawa-Sakata(MNS) 
form\cite{mns} for the 
mixing matrix then we can identify the element $U_{e3}$ with the 
mixing angle $\sin \theta_{13}$ which is the relevant angle for 
the CHOOZ experiment. 
Since $\sin \theta_{13}=\epsilon\cos 2\theta$ we get $\sin \theta_{13} < 0.07$ 
or in other words $\tan^2\theta_{13}<0.005$ 
from the allowed values of $\theta$ and $\epsilon$ which is well within 
the CHOOZ bound \cite{chooz}. 
Thus the bimaximal solution of Zee model is seen to be consistent with 
the CHOOZ experiment.

We next examine the range of values for $\Delta m_{atm}^2$ 
and $\sin^22\theta_{atm}$ allowed at both at 90\% and 99\% C.L. 
from the analysis of
the latest SK atmospheric neutrino data \cite{flatm} and find the
corresponding range of $\epsilon$ for $\Delta m_{\odot}^2$ in the solar
range at 90\% and 99\% levels. This can be done using
Eqn. (\ref{delmsq}). We show this range of $\epsilon$ as a
function of $\Delta m_{\odot}^2$ in fig. \ref{fig1} only at 99\% C.L.. In the 
left hand panel we show the range of $\epsilon$ required to generate the 
$\Delta
m_{\odot}^2$ splitting in the LMA region while the right hand panel gives
the corresponding splitting in the LOW region. The range of $\Delta
m_{\odot}^2$ shown are allowed at 99\% C.L. from the global
analysis of the most recent analysis of the solar data including SNO
\cite{bcgk}. We note that while our approximation of treating $\epsilon$
perturbatively is correct in the LOW region of the solar neutrino solution
it may not be fully justified for the higher $\Delta m_{\odot}^2$ allowed
in the LMA zone as the value of $\epsilon$ is quite large. 
Let us next look at the values of the solar mixing angle predicted 
by the allowed range of $\epsilon$ shown in fig. \ref{fig1}.
This can be done using Eqn. (\ref{permix}) where $U_{e2}$ element
depends on $\epsilon$ when we turn on our $L_e - L_\mu -L_\tau$
breaking perturbation. 
We can identify 
$\cos\theta_{\odot}\cos\theta_{13} = U_{e1}$ and 
$\sin\theta_{\odot}\cos\theta_{13} = U_{e2}$. 
In fig. \ref{fig2} the red horizontal bars show the 
range of values for the predicted solar mixing angle 
inputing
mass and
mixing required to explain the SK atmospheric neutrino data at 99\%
C.L. plus necessary $\epsilon$ required for the solar neutrino
problem at 99\% C.L.. 
Dotted lines are the allowed areas in the 
$\Delta m_{\odot}^2- \tan^2\theta_{\odot}$ plane at 99\% C.L. 
from global analysis of solar neutrino data including 
SNO \cite{bcgk}. The figure clearly shows that in the LMA zone 
there is no overlap between the values
of $\Delta m_{\odot}^2$ and $\tan^2\theta_{\odot}$ predicted by the Zee
model and those that are allowed by the current data at 99\% C.L.. In the
LOW region however the Zee model is found to be consistent with the
experiments at 99\% C.L. in the analysis of \cite{bcgk}. 
We see that the allowed areas in the 
VO region of \cite{bcgk} is also in conflict 
with bimaximal solution of the Zee model. 
Thus we infer it to be an interesting
hint of new physics of Zee type which can produce the
LOW solution but cannot produce LMA or VO solutions
for the solar neutrino problem and simultaneously
can produce maximal $\nu_\mu \leftrightarrow \nu_\tau$
mixing to produce the atmospheric neutrino oscillations.
In fig. \ref{fig2} dashed lines are the 99\% C.L. allowed areas 
obtained by Bahcall {\it et al.} in \cite{bcc} (the contours shown 
have been read\footnote{For exact 
values the reader should refer to \cite{bcc}.} 
from the fig. 1 of \cite{bcc}). 
However in \cite{bcc} Bahcall {\it et al.} use 
a different data analysis technique than that used in \cite{bcgk,others}.
It is clear that Zee model is consistent with the analysis 
of \cite{bcc} at 99\% C.L. not only in the LOW solution but also 
in the LMA region. In fact if one looks at the fig. 1 
of \cite{bcc}, LOW is compatible 
with Zee model at the 90\% C.L.. So we see that
the compatibility of Zee model in the LOW
region at 99\% C.L. is actually a conservative estimate. 
We notice the same trend in all the different analysis and 
conclude that the LOW solution is more consistent with maximal mixing
in the light of present data. We emphasize that
even though the LMA is the `best-fit' solution from experimental data,
the LOW solution can give a comparable description of the 
experimental data \cite{bcgk,others,bcc} and this signifies that the 
minimal Zee model can give an explanation of the solar 
neutrino data when the amount of $L_e - L_\mu - L_\tau$ breaking is
such that $\Delta m^2_\odot$ falls in the LOW zone. 

Credibility of bimaximal solution in Zee model depends obviously on the 
compatibility of the maximal mixing solution with the solar neutrino 
data. The maximal mixing solution is at variance with the global data 
mainly due to the fact that the Cl data is about 
$2\sigma$ lower compared to the rate predicted for the Cl experiment 
by the maximal mixing solution (see fig. 7 of \cite{ggmax} and 
fig. 2 of \cite{eind2}). The Ga data 
in the LMA region is higher compared to the predicted maximal mixing rate, 
though in the LOW region the agreement improves due to earth 
matter effects. 
The rate expected
in the Borexino experiment is $\approx 0.62$ if one has maximal
mixing with $\Delta m_{\odot}^2$ in the LMA region, while 
for the LOW solution the rate expected is a little higher.
Borexino expects to see significant earth
regeneration effect in the LOW region, resulting in more events during
night than during day. However for the LMA solution there will not 
be much day-night asymmetry. 
Thus Borexino has good sensitivity in $\Delta m^2_{\odot}$ and 
holds the potential to distinguish between the LMA and LOW solutions. 
SNO on the other hand is not very sensitive to
$\Delta m^2_\odot$
since it expects to see almost the same neutrino
rate for both the LMA and the LOW regions. However it is very sensitive to
the value of the mixing angle $\theta_\odot$. The ratio of charged 
to neutral current (CC/NC) events in SNO is the best variable to 
look at. In future the entire LMA region will also be scanned 
by the KamLand reactor experiment which is expected to
observe the actual oscillations.

In conclusion, Zee model is so predictive because it has only three 
real parameters from which one can calculate three masses and three 
mixing angles, thus it has three predictions. 
Bimaximal mixing solution demands an approximate $L_e - L_\mu -L_\tau$
symmetry. Question is how badly this flavor symmetry is broken ?
The value of $\Delta m_{\odot}^2$ parameterizes the extent to which 
this flavor symmetry is broken. In the light of present data the minimal 
version of Zee model is in better agreement with the 
LOW solution to the solar neutrino problem than the LMA solution or VO 
solution. This is independent of the fact that LMA is the current 
best-fit. But since the LOW solution gives an acceptable fit to the 
solar data and since the Zee model is consistent with the LOW solution of 
solar neutrino problem as well as atmospheric neutrino anomaly, it is at 
present a viable model. Furthermore the 13 element of the mixing matrix is
fully consistent with CHOOZ experiment. With a wealth of data awaited 
from the future solar neutrino experiments and more data awaited from 
SNO one may hope to further test the compatibility of bimaximal 
solution of Zee model.   

We thank Ambar Ghosal, Yoshio Koide, Ernest Ma and Tadashi Yoshikawa for
communications on Zee model.

\newpage

\begin{figure}
\begin{center}
\begin{tabular}{c}
\epsfysize=12cm \epsfxsize=12cm \hfil \epsfbox{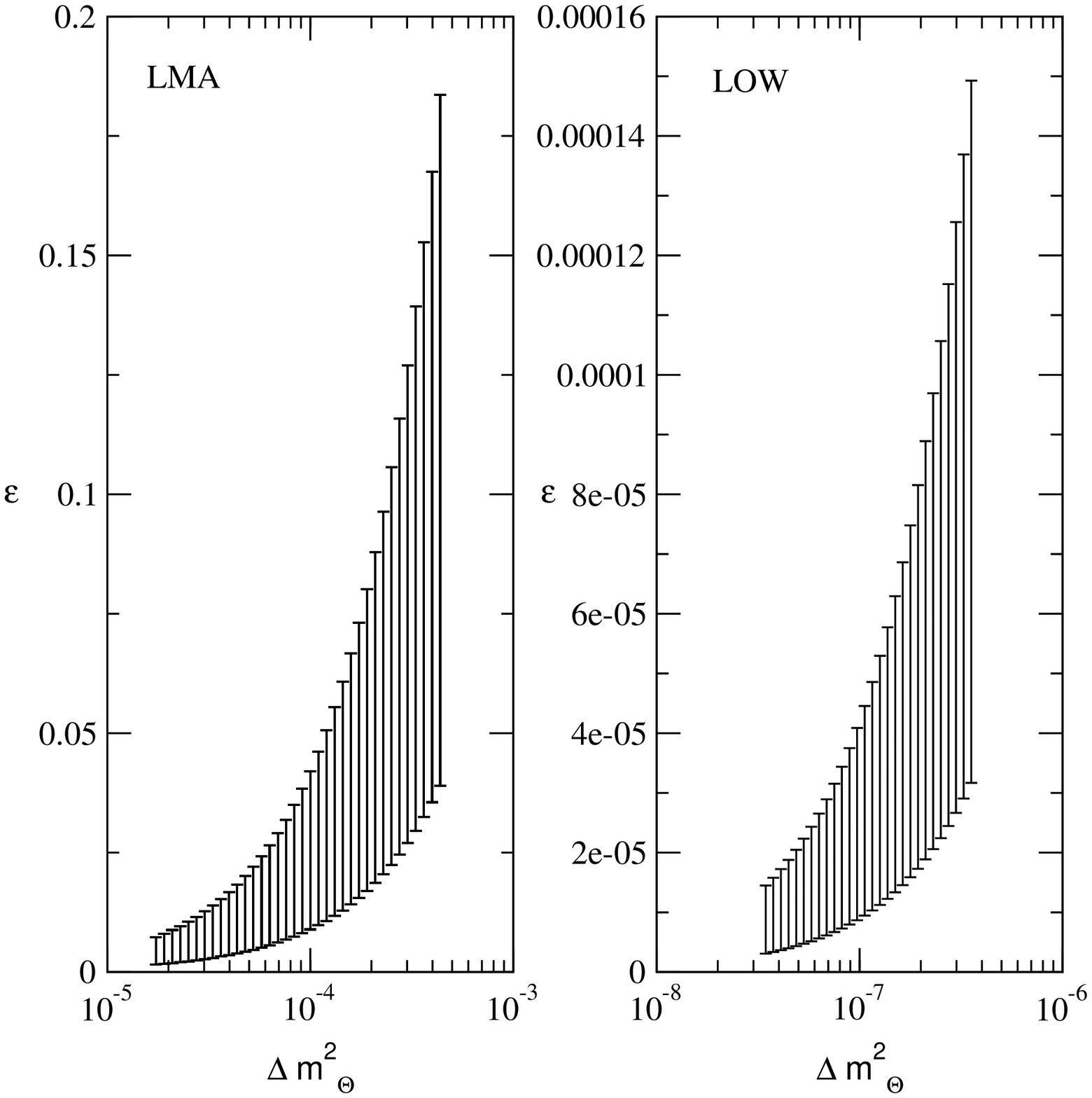} \hfil
\end{tabular}
\caption{ 
 The range of $\epsilon$ as a function of $\Delta m_{\odot}^2$
corresponding to the 99\% C.L. range of allowed values of
$\Delta m_{atm}^2$ and $\sin^2 2\theta_{atm}$ from \cite{flatm}. The
$\Delta m_{\odot}^2$ shown corresponds to the 99\% C.L.
allowed range for LMA (left hand panel) and LOW (right hand panel)
solutions from \cite{bcgk}.}
\label{fig1}
\end{center}
\end{figure}

\begin{figure}
\begin{center}
\begin{tabular}{c}
\epsfysize=12cm \epsfxsize=12cm \hfil \epsfbox{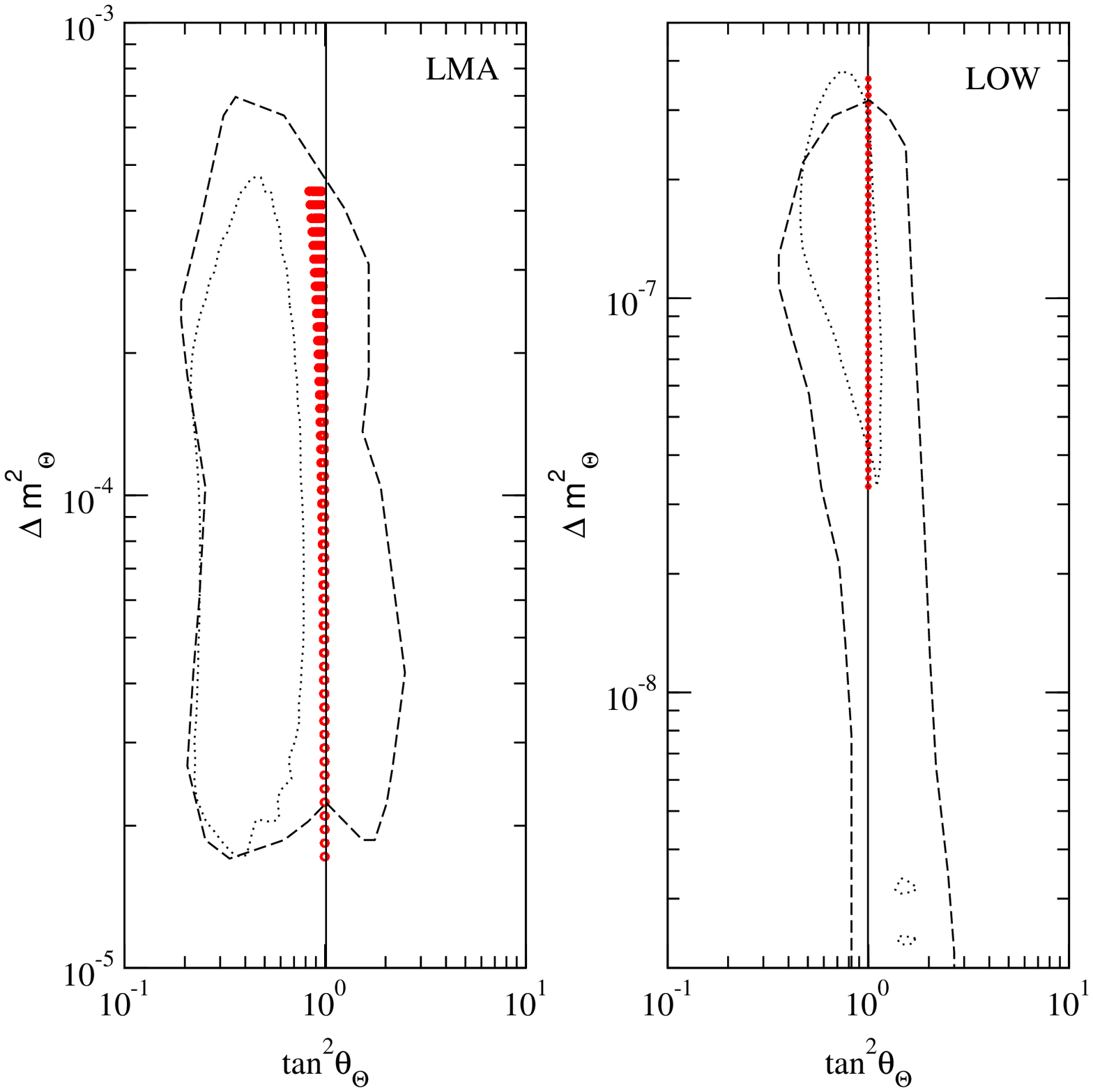} \hfil
\end{tabular}
\caption{
Comparison of the Zee model predictions with the
allowed parameter values from the current experimental data.
Values of $\tan^2\theta_{\odot}$ and $\Delta m_{\odot}^2$
predicted by the Zee model are shown by the red horizontal
errorbars. The dotted lines give the 99\% C.L. allowed areas in the
LMA (left hand panel) and LOW (right hand panel) regions from the
analysis of \cite{bcgk}. The dashed lines give corresponding
allowed zones at 99\% C.L. from the analysis of Bahcall {\it et al}
\cite{bcc}.}
\label{fig2}
\end{center}
\end{figure}


\begin{thebibliography}{99}                                                

\bibitem{zee}
A. Zee, Phys. Lett. {\bf B93}, 389 (1980); Phys. Lett. {\bf B 161}, 141 (1985);
L. Wolfenstein, Nucl. Phys. {\bf B175}, 93 (1980); S. T. Petcov, Phys. Lett. 
{\bf B115}, 401 (1982). 

\bibitem{smirnov} 
A. Yu. Smirnov and M. Tanimoto, Phys. Rev. {\bf D55}, 1665 (1997)

\bibitem{jarlskog} 
C. Jarlskog, M. Matsuda, S. Skadhauge, M. Tanimoto, Phys. Lett. {\bf B449}, 240 (1999)

\bibitem{petcov} S.T. Petcov, Phys. Lett. {\bf B110}, 245 (1982). 

\bibitem{frampton} P. H. Frampton, S. L. Glashow, 
Phys. Lett. {\bf B461}, 95 (1999)


\bibitem{newpapers}
E. Mitsuda, K. Sasaki, Phys. Lett. {\bf B516} 47 (2001);
A. Ghosal, Y. Koide, H. Fusaoka, Phys. Rev. {\bf D64}, 053012 (2001);
K.R.S. Balaji, W. Grimus, T. Schwetz, Phys. Lett. {\bf B508} 301 (2001);
Y. Koide, A. Ghosal, Phys. Rev. {\bf D63}, 037301 (2001); 
N. Gaur, A. Ghosal, E.  Ma, P. Roy, Phys. Rev. {\bf D58} 071301 (1998); 
S. Kanemura, T. Kasai, G. L. Lin, Y. Okada, J.-J. Tseng, C.P. Yuan
Phys. Rev. {\bf D64}, 053007 (2001);
K. Cheung, O. C.W. Kong, Phys. Rev. {\bf D61}, 113012 (2000);
A. Yu. Smirnov, Z-j Tao, Nucl. Phys. {\bf B426}, 415 (1994). 

\bibitem{koide}Y. Koide, Phys. Rev. {\bf D64}, 077301 (2001). 

\bibitem{frampton1} P. H. Frampton, M. C. Oh, T. Yoshikawa, 
e-Print Archive: hep-ph/0110300


\bibitem{mns}
B. Pontecorvo, Zh. Eksp. Teor. Fiz. {\bf 33}, 549 (1957);
{\bf 34}, 247 (1958);
Z. Maki, N. Nakagawa, S. Sakata, Prog. Theor. Phys.
{\bf 28}, 870 (1962).

\bibitem{chooz}  M. Appolonio \etal, Phys. Lett. {\bf B466},
415 (1999); Phys. Lett. {\bf B420}, 397 (1998).

\bibitem{flatm}  G.L.Fogli, E.Lisi, A. Marrone, 
e-Print Archive: hep-ph/0110089. 

\bibitem{bcgk} A. Bandyopadhyay, S. Choubey, S. Goswami , K. Kar,
 Phys. Lett. {\bf B519}, 83 (2001). 

\bibitem{others}  G.L. Fogli, E. Lisi, D. Montanino, A. Palazzo,
Phys. Rev. {\bf D64}, 093007 (2001); 
P.I. Krastev and A.Yu. Smirnov, e-Print Archive: hep-ph/0108177; 
M.V. Garzelli and C. Giunti, e-Print Archive: hep-ph/0108191.

\bibitem{bcc}J.N. Bahcall, M.C. Gonzalez-Garcia, C. Pana-Garay,
JHEP {\bf 0108}, 014 (2001). 

\bibitem{ggmax} M.C. Gonzalez-Garcia,  C. Pana-Garay, Y. Nir 
, A.Yu. Smirnov,  Phys. Rev. {\bf D63}, 013007 (2001). 

\bibitem{eind2} S. Choubey, S. Goswami, D.P. Roy,
e-Print Archive: hep-ph/0109017, to appear in Phys. Rev. D.


\end{thebibliography}
\end{document}